# Excellent Thermoelectric Performances of Pressure Synthesized ZnSe$_2$

Tiantian Jia[1,2,3], Jesús Carrete[3], Zhenzhen Feng[1,2], Shuping Guo[1,2], Yongsheng Zhang[1,2,*], and Georg K.H. Madsen[3,*]

1) Key laboratory of Materials Physics, Institute of Solid State Physics, Chinese Academy of Sciences, 230031 Hefei, P. R. China
2) University of Science and Technology of China, 230026 Hefei, P. R. China
3) Institute of Materials Chemistry, TU Wien, A-1060 Vienna, Austria

Corresponding author:
yshzhang@theory.issp.ac.cn (Y. Zhang)
georg.madsen@tuwien.ac.at (G.K.H. Madsen)

## ABSTRACT

We calculate the lattice thermal conductivities of the pyrite-type ZnSe$_2$ at pressures of 0 and 10 GPa using the linearized phonon Boltzmann transport equation. We obtain a very low value [0.69 W/(m·K) at room temperature at 0 GPa], comparable to the best thermoelectric materials. The vibrational spectrum is characterized by the isolated high-frequency optical phonon modes due to the stretching of Se-Se dimers and low-frequency optical phonon modes due to the rotation of Zn atoms around these dimers. The low-frequency optical phonon modes are characterized by a strong anharmonicity and will substantially increase the three-phonon scattering space which suppress the thermal conductivity. Interestingly, two transverse acoustic phonon modes with similar frequencies and wave vectors have very different degrees of anharmonicity depending on their polarization. We relate this to the low thermal conductivity and show that the anharmonicities of the transverse acoustic phonon modes are connected to the corresponding change in the pyrite parameter, which can be interpreted as a descriptor for the local volume change. To determine the thermoelectric performance of ZnSe$_2$, we also investigate its electrical transport properties. The results show that both p-type or n-type ZnSe$_2$ can show promising electrical transport properties. We trace this back to the complex energy isosurfaces of both valence and conduction bands. The low thermal conductivities and promising electrical transport properties lead to a large thermoelectric figure of merit of ZnSe$_2$ for both p-type and n-type doping.

# INTRODUCTION

Thermoelectric (TE) materials can generate electricity from waste heat. They therefore have the potential to play an important role in solving the current energy and environmental crisis. The thermoelectric conversion efficiency is determined by the dimensionless thermoelectric figure of merit $ZT = S^2\sigma T/\kappa$, where $T$ is the absolute temperature, $S$ is the Seebeck coefficient, σ is the electrical conductivity and $\kappa$ is the thermal conductivity comprising both charge carrier ($\kappa_e$) and lattice ($\kappa_l$) contributions. The larger the $ZT$ value, the greater is the maximum achievable thermoelectric conversion efficiency. Based on the definition of $ZT$, the promising thermoelectric materials benefit from a large power factor ($PF = S^2\sigma$) and a low $\kappa$. However, it is very difficult to simultaneously achieve large electrical transport coefficients ($S$ and $\sigma$) and low thermal conductivity in a compound, since $S$, σ and $\kappa_e$ are carrier concentration dependent and therefore correlated with each other.[1] Furthermore, $\kappa_l$ is only partially independent of σ because the same factors enhancing phonon scattering can also impede charge carrier transport.

With the introduction of the Phonon-Glass-Electron-Crystal[2] concept, many mechanisms have been tried to achieve high $PF$ and low $\kappa_l$: Band-structure engineering (including the multiple-band alignment, resonant states and energy filtering mechanisms, et al.)[3-6] has been used to improve the $PF$. Multiscale phonon scattering approaches, such as the nanostructure engineering, low-dimensional designs, and strong anharmonicity caused by lone pairs of electrons or liquid-like atoms, et al.,[7-11] have been proposed as ways to decrease $\kappa_l$. However, the currently studied TE materials still have relatively low $ZT$ values ($ZT < 3$)[12], resulting in low energy conversion efficiency, or contain toxic or non-abundant chemical elements (such as Pb and Te). This limits the commercial application of TE devices and makes the quest for new TE materials with good thermoelectric performances still ongoing.

Some of us have recently investigated the thermoelectric properties of 243 known binary semiconductor chalcogenides using high-throughput computations, and predicted some new high-performance TE materials based on their electrical and thermal descriptors.[13] Among them, ZnSe$_2$ possesses both promising thermal and electrical transport properties, as well as having a simple crystal structure and being made up of non-toxic and abundant elements. The dichalcogenide (ZnSe$_2$) was successfully synthesized at high pressure (6.5 GPa) and high temperature (600 – 800 ºC), in 1968, and found to be (meta-)stable at room temperature and ambient pressure.[14] Therefore, ZnSe$_2$ deserves further attention both to clarify its transport properties using more accurate quantitative predictive methods and to understand the physical mechanisms behind its promising thermoelectric behavior.

In this work, using first-principles calculations, we study the phase stability of ZnSe$_2$ at different pressures. We find that ZnSe$_2$ is mechanically stable both under and at ambient pressure. Using the full linearized phonon Boltzmann transport equation (BTE), we calculate the lattice thermal conductivities of ZnSe$_2$ at 0 and 10 GPa. Our results show that ZnSe$_2$ has a very low lattice thermal conductivity. This is attributed partly to low-frequency optical modes stemming from the rotational motion of Zn atoms around Se-Se dimers. These can substantially increase the three-phonon scattering space, and are also characterized by a strong anharmonicity. Moreover, a strong anharmonicity in the acoustic phonon region is also pointed out and attributed to the transverse acoustic phonon mode with a strong local change of the pyrite parameter. Using the acoustic phonon scattering time approach, we further calculate the electrical transport properties of ZnSe$_2$ in the framework of the electron BTE. We find that both p-type and n-type ZnSe$_2$ exhibit promising electrical transport properties. The performance relates to the complex energy isosurfaces of both valence and conduction bands. Combing the low thermal conductivities and promising electrical transport properties, we find that ZnSe$_2$ is an excellent thermoelectric material for both p-type and n-type doping.

## COMPUTATIONAL METHODOLOGIES

### I. Density-functional theory calculations

The electronic structure of ZnSe$_2$ is calculated utilizing density functional theory (DFT) as implemented in the Vienna Ab initio Simulation Package (VASP)[15] with the projector augmented wave (PAW)[16] method. The electronic exchange and correlation energy are accounted for the generalized gradient approximation of Perdew, Burke, and Ernzerhof (GGA-PBE)[17]. The energy cutoff for the plane-wave expansion is 420 eV. The geometry is relaxed until the change in total energy is less than $10^{-5}$ eV and the force components of each atom are below 0.01 eV/Å. The Brillouin zone is sampled using a (5 × 5 × 5) Monkhorst-Pack k-point mesh[18]. Since the GGA is well-known for underestimating the band gaps of semiconductors and insulators, we further carry out calculations with the Heyd–Scuseria–Ernzerhof functional (HSE06)[19,20] to achieve the more accurate band gap of $ZnSe_2$.

The phonon vibrational properties are calculated using the finite displacement method (as implemented in the PHONOPY software package[21]). To obtain reliable phonon dispersions, we need a highly optimized crystal structure. Thus, a (6 × 6 × 6) k-point mesh, a $10^{-8}$ eV energy convergence criterion and a $10^{-4}$ eV/Å force convergence criterion are adopted to further optimize the crystal structure of $ZnSe_2$, and a (2 × 2 × 2) supercell with 96 atoms is used. In the quasiharmonic DFT phonon calculations, the crystal volumes are isotropically changed by ±3% based on the DFT relaxed volume.

The lattice thermal conductivities ($\kappa_l$) are calculated by solving the full linearized phonon BTE as implemented in the almaBTE code[22]. In the almaBTE method, the Born effective charges and the dielectric tensor are calculated using VASP. To determine the amplitudes of the three-phonon processes, the third-order interatomic force constants (IFCs) are computed using the thirdorder.py code[23]. We consider interatomic interactions up to the 7$^{th}$ coordination shell. A (10 × 10 × 10) Γ-centered grid is used to solve the phonon BTE.

## II.   Electrical transport property calculations

The electrical transport properties ($S$, $\sigma/\tau$ and $\kappa_e$) are calculated using the BoltzTraP. code[24], which is based on the electron BTE. A highly dense (35 × 35 × 35) k-point mesh is used to obtain accurate electrical transport properties. Since the wavelength of a thermal electron is comparable with the lattice constant, the acoustic phonon scattering is typically the dominant carrier scattering mechanism.[25] We therefore model the carrier relaxation time by considering the acoustic phonon scattering mechanisms and using the single parabolic band model.[25] In the model, the carrier relaxation time of a single energy valley ($\tau_i$) is given by the deformation potential approach as

$$\tau_i = \frac{2^{\frac{1}{2}}\pi\hbar^4 \rho v_{LA}^2}{3E_{di}^2 (m_{bi}^* k_B T)^{\frac{3}{2}}} \frac{F_0(\eta)}{F_{\frac{1}{2}}(\eta)}$$

$$F_x(\eta) = \int_0^\infty \frac{E^x}{1+exp(E-\eta)} dE, \quad \eta = \frac{\mu}{k_B T} \qquad (1)$$

where $\hbar$ is the reduced Planck constant, $k_B$ is the Boltzmann constant, $\rho$ is the mass density, $v_{LA}$ is the longitudinal acoustic (LA) phonon velocity, $\eta$ is the reduced chemical potential, $m_{bi}^*$ is the single valley density of states (DOS) effective mass and $E_{di}$ is the deformation potential of the $i^{th}$ valley. For a degenerate semiconductor, the total carrier relaxation time ($\tau$) can be evaluated from[26]

$$\frac{1}{\tau} = \sum_i \frac{D_i}{\tau_i} \quad (2)$$

where, $D_i$ is the symmetry degeneracy of the $i^{th}$ valley. $v_{LA}$ can be calculated from the weighted averages of the particular group velocity ($v_{LA}^{\alpha\beta}$) along ($\alpha, \beta$) wave vector as

$$v_{LA} = \frac{1}{\sum D_{\alpha\beta}} \sum_{\alpha\beta} D_{\alpha\beta} v_{LA}^{\alpha\beta} \quad (3)$$

$$v_{LA}^{\alpha\beta} = \frac{\partial \omega_{LA}^{\alpha\beta}}{\partial k_{\alpha\beta}}, (\alpha, \beta = x, y, z) \quad (4)$$

where, $D_{\alpha\beta}$ is the symmetry degeneracy of the ($\alpha, \beta$) direction, $k_{\alpha\beta}$ is the reciprocal space wave vector and $\omega$ is the phonon frequency. $m_b^*$ can be obtained from the geometric mean effective mass of a single ellipsoid energy valley, as[27]

$$m_b^* = \left(m_\parallel^* m_\perp^{*2}\right)^{\frac{1}{3}} \quad (5)$$

where $m_\parallel^*$ and $m_\perp^*$ are the principal effective masses parallel and perpendicular to the long axis of the elliptic energy valley, respectively. They can be calculated from the band structures as

$$m_{\parallel/\perp} = \hbar^2 \left[\frac{\partial^2 E(k)}{\partial k_{\parallel/\perp}^2}\right]^{-1} \quad (6)$$

$E_d$ characterizes the changes in electronic energies with volume, which is defined as

$$E_d = \Delta E / \left(\frac{\Delta V}{V}\right) \quad (7)$$

where $\Delta E$ is energy change of the band extrema of the single energy valley [the valence band maximum (VBM) or the conduction band minimum (CBM)] with the volume dilation $\frac{\Delta V}{V}$. The deformation potential approach has been successfully applied to predict $ZT$ values in a number of investigations of the thermoelectric materials.[28-30]

## RESULTS AND DISCUSS

### I. Crystal structure and stability of ZnSe$_2$

The dichalcogenide ZnSe$_2$ with a pyrite-type crystal structure (Fig. 1a) was synthesized at 6.5 GPa in 1968 and found to be (meta-)stable at ambient pressure.[14] The pyrite-type ZnSe$_2$ is a cubic structure (space group Pa$\bar{3}$) with Wyckoff position 4a(0,0,0) for Zn and 8c($u,u,u$) for Se, where $u$ is the pyrite parameter. In the structure, each Zn has six Se nearest neighbors and each Se has three Zn and one Se nearest neighbors. In this work, our relaxed lattice constant (6.366 Å) at 0 GPa is larger than the experimental (6.293 Å), as is typically found for the PBE functional. And the calculated bond lengths of Zn-Se ($L_{Zn-Se}$) and Se-Se ($L_{Se-Se}$) are 2.68 and 2.40 Å at 0 GPa.

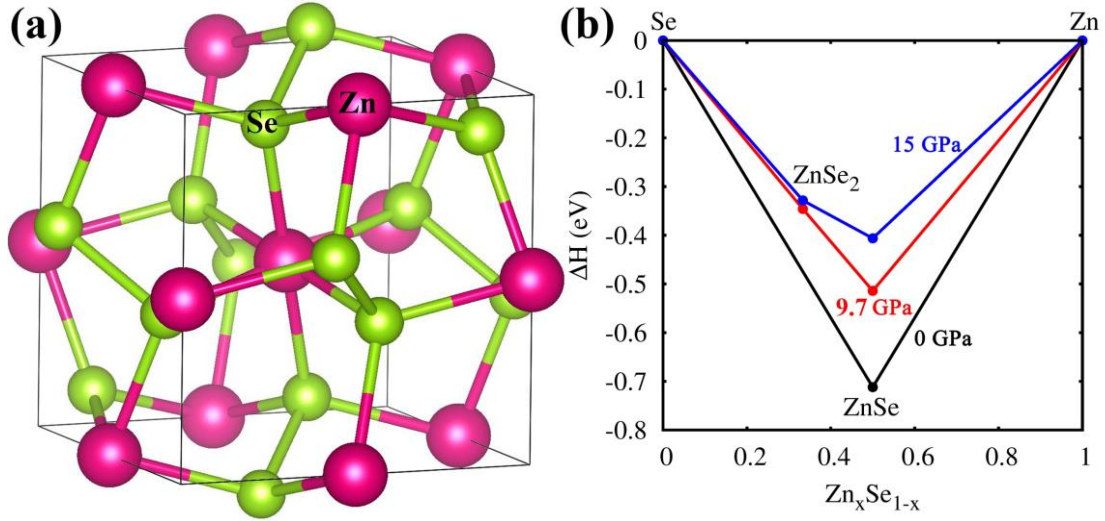

Figure 1. (a) The pyrite-type crystal structures of ZnSe$_2$ (space group Pa$\bar{3}$). (b) DFT calculated pressure-composition phase diagrams of Zn$_x$Se$_{1-x}$ at 0 (black), 9.7 (red) and 15(blue) GPa, respectively.

To check the thermodynamic stability of ZnSe$_2$ under pressures, we calculate the formation enthalpy [$\Delta H(P)$] of Zn$_x$Se$_{1-x}$ under different pressures,

$$\Delta H(P) = H(Zn_xSe_{1-x}) - xH(Zn) - (1-x)H(Se) \qquad (8)$$

where x is the composition of Zn in the Zn-Se compound, and $H(= E + PV)$ is the pressure dependent enthalpy at different pressures ($P$) and volumes ($V$). We calculate the formation enthalpies of Se, ZnSe, ZnSe$_2$ and Zn for different phases[31] in the pressure range from 0 to 15 GPa and setup the pressure-dependent convex hulls of Zn-Se (Fig. 1b and Fig. S1a in detail). At 0 GPa, we find that ZnSe (the zincblende phase) is the thermodynamically stable. However, the formation enthalpy of ZnSe$_2$ is only ~0.14 eV above the convex hull, which could indicate that ZnSe$_2$ is a metastable compound. Additionally, due to pressure having a stronger influence on the formation enthalpy in ZnSe than in ZnSe$_2$: increasing pressure from 0 to 15 GPa, ΔH increases ~0.306 eV in ZnSe, while only ~0.001 eV in ZnSe$_2$, as shown in Fig. S1b in Supplementary Materials, ZnSe$_2$ becomes a thermodynamically stable compound at pressure higher than 9.7 GPa. The predicted pressure (9.7 GPa) is in reasonable agreement with the synthesis pressure (6.5 GPa)[14]. This slight pressure difference between theoretical calculation and experimental measurement may be due to the fact that ZnSe$_2$ was synthesized under the combination of pressure and temperature, while in our theoretical study, the calculated formation enthalpy only considers the influence of pressure not temperature.

Due to the theoretically predicted pressure induced thermodynamically stability of ZnSe$_2$ being 9.7 GPa, we calculate the geometric, vibrational and electrical properties of ZnSe$_2$ at 10 GPa. At 10 GPa, the lattice constant (*a*) of ZnSe$_2$ shrinks to 6.102 Å, and the calculated $L_{Zn-Se}$ and $L_{Se-Se}$ are compressed to 2.56 and 2.36 Å. Compared to the calculated interatomic bond lengths at 0 GPa, it is clear that $L_{Zn-Se}$ (0.12 Å) has changed substantially more than $L_{Se-Se}$ (0.04 Å).

In order to study the mechanical stability of ZnSe$_2$, we calculate the phonon dispersions of ZnSe$_2$ at 0 (Fig. 2a) and 10 GPa (Fig. 2c). From Fig. 2, we find that all phonon frequencies are real, which indicates that ZnSe$_2$ is mechanically stable both under and at ambient pressure. The stable phonon behavior of ZnSe$_2$ at 0 GPa underlines that even though the compound is synthesized at high pressure, it should be metastable at ambient pressure. At 0 GPa, the phonon density of states (PDOS) per atom of ZnSe$_2$ is shown in Fig. 2b. The most striking thing is the presence of the localized phonon modes at ω ≈ 250 cm$^{-1}$, which are almost entirely made up of Se vibrations.

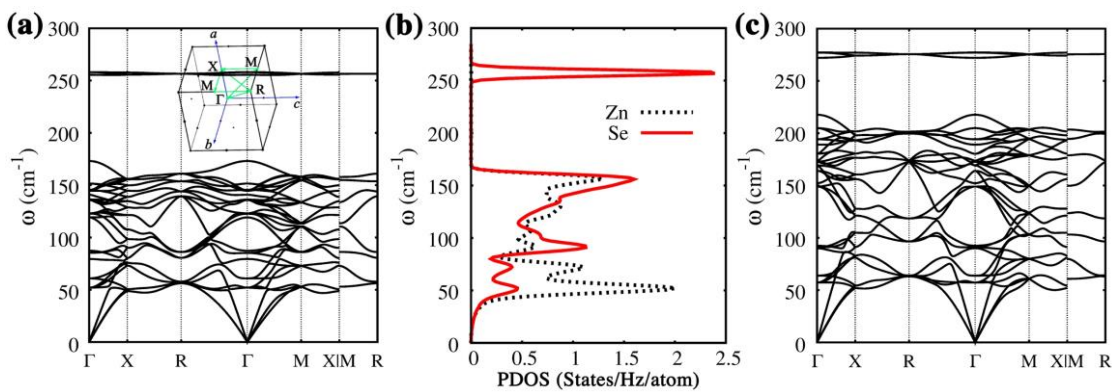

Figure 2. Phonon dispersions of ZnSe$_2$ at 0 (a) and 10 GPa (c). The first Brillouin zones of ZnSe$_2$ with high-symmetry points is illustrated as an insert. (b) The corresponding phonon density of states (PDOS) per atom of ZnSe$_2$ at 0 GPa.

To understand the origin of these high-frequency phonon modes and further illustrate the bonding situation in ZnSe$_2$, we calculate the electron localization function (ELF)[32], as shown in Fig. 3a and 3b. The ELF measures the extent of spatial localization of the reference electron located at a given position, with values ranging from 0 to 1. ELF = 0.5 corresponds to the electron gas, while ELF = 1 corresponds to perfect localization (as found in regions with covalent bonds, core shells, and lone pairs). Fig. 3a and 3b show the calculated three dimensional (3D) and two dimensional (2D) [(001) plane] ELF for ZnSe$_2$, respectively. From Fig. 3a, we notice that there is a dumbbell ELF shape around the Se-Se bond. The dumbbell shape can be interpreted as the presence of Se-Se dimers connected by a strong covalent bond. A detailed analysis of the phonon dispersions reveals that the stretches of Se atoms along Se-Se dimers contribute the high-frequency localized phonon modes at $\omega \approx 250$ cm$^{-1}$ (as shown in Fig. S2a in Supplementary Materials).

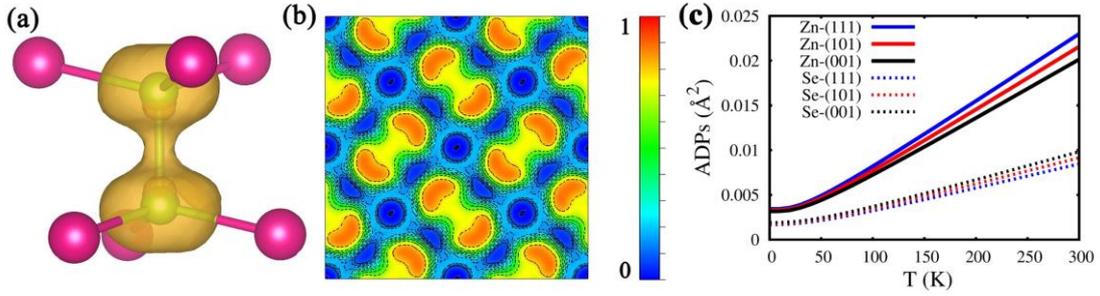

Figure 3. (a) 3D electron localization function (ELF) of ZnSe$_2$ at 0 GPa. The ELF value of the isosurface is 0.7. (b) Projected 2D ELF in (001) plane of ZnSe$_2$ at 0 GPa. (c) Calculated atomic displacement parameters (ADPs) of Zn and Se atoms along different directions.

To further understand the vibrational properties of different atoms, we calculate atomic displacement parameters (ADPs, the mean squared displacement amplitude of the atom relative to its equilibrium position) of the Zn and Se atoms along different directions, as shown are in Fig. 3c. The ADPs of Zn atoms are much larger than those of Se atoms, indicating that Zn atoms are more loosely bound. This agrees well with the change of $L_{\text{Zn}-\text{Se}}$ being greater than that of $L_{\text{Se}-\text{Se}}$ at 10 GPa and the low-frequency optical phonons (40 cm$^{-1} < \omega <$ 80 cm$^{-1}$) having larger Zn than Se contributions, Fig. 2b, despite the lower mass of Zn.

## II. Low lattice thermal conductivities

In our previous high-throughput work[13], we predicted that ZnSe$_2$ could possess a low lattice thermal conductivity ($\kappa_l$). The prediction was based on a strong dependence of the elastic properties on volume, which could indicate a strong anharmonicity or, more specifically, a large Grüneisen parameter ($\gamma$). In this work, we explicitly calculate the thermal conductivity using the full linearized phonon BTE coupled with DFT, which is well-known for quantitative predictive power[33-37]. Specifically, we calculate $\kappa_l$ at both 0 and 10 GPa, as shown in Fig. 4, and find that ZnSe$_2$ has a very low lattice thermal conductivity: $\kappa_l$ is equal to 0.69 and 1.98 W/(m·K) at 300 K at 0 and 10 GPa, respectively. These values are comparable to the know low-thermal conductivity compounds for thermoelectric applications, such as SnSe[8] and PbTe[38] [the measured lattice thermal conductivities of the two compounds are 0.45-0.7 and 2.4 W/(m·K) at room temperature, respectively].

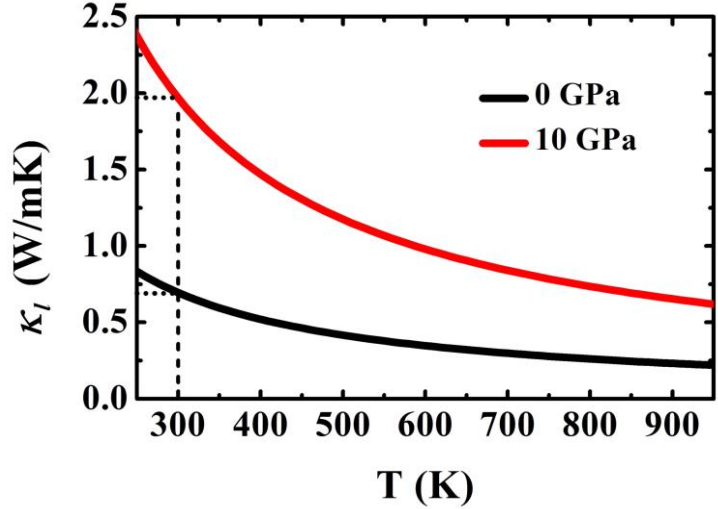

Figure 4. Calculated $\kappa_l$ as a function of temperature at 0 and 10 GPa, respectively.

To verify the applicability of the phonon BTE to a situation with such low thermal conductivities, we calculate the minimum lattice thermal conductivity ($\kappa_{min}$) from

$$\kappa_{min} = \frac{1}{3}\sum_q C_q v_q \frac{v_q}{\omega_q} \qquad (7)$$

where $C_q$ [J/(K·m$^3$)] is the contribution of each phonon to the volumetric heat capacity. This value approximates the thermal conductivity that the system would display if every phonon has a mean free path that put it at the Ioffe-Regel crossover, below which the phonon-based description is no longer useful.[39] The calculated $\kappa_{min}$, 0.09 and 0.10 W/(m·K) at 300 K at 0 and 10 GPa, are well below $\kappa_l$ in Fig. 4, lending support to our choice of the formalism.

## A. Grüneisen parameters

The mode dependent Grüneisen parameters ($\gamma_{iq} = -\frac{V}{\omega_{iq}} \frac{\partial \omega_{iq}}{\partial V}$) can be viewed as weighted averages of the 3$^{rd}$ order force constants which enter the expression for the scattering rates.[40] Several previous studies have linked a large Grüneisen parameter with a low thermal conductivity.[41-43] To investigate the origins of the low $\kappa_l$ of ZnSe$_2$, we evaluate $\gamma_{iq}$ on a mode-by-mode basis by calculating the change of the phonon vibrational properties ($\omega_{iq}$) when the volume of system is isotropically changed by ±3% with respect to the DFT relaxed volume at 0 GPa. The calculated $\gamma_{iq}$ at different phonon modes are shown in Fig. 5a. On Fig. 5a, we find that many phonon modes have large anharmonicities ($\gamma > 2$) and the maximum Grüneisen parameter located at one of the transverse acoustic phonon branches (TA) is 4.6, which is comparable to the well-known SnSe compound with large anharmonicities ($\gamma_a = 4.1$, $\gamma_b = 2.1$ and $\gamma_c = 2.3$ along the $\vec{a}$, $\vec{b}$ and $\vec{c}$ three directions, respectively)[44].

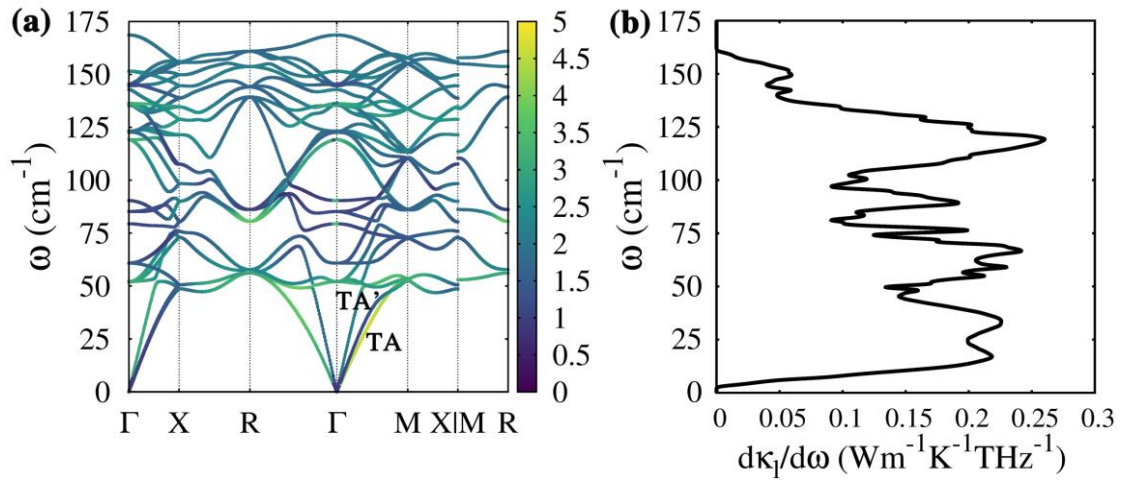

Figure 5. (a) The phonon band structures and the corresponding distribution of $\gamma_{iq}$ at 0 GPa. The color in (a) denotes the values of $\gamma_{iq}$. (b) The $\kappa_l$ contribution with respect to phonon frequency at 300 K.

Through computing the $\kappa_l$ contribution for ZnSe$_2$ with respect to phonon frequency at 300 K (Fig. 5b), we find that the high-frequency phonons contributed by Se-Se dimers ($\omega \sim 250$ cm$^{-1}$) make almost no contribution to the thermal conductivity. Instead, the thermal conductivity has contributions from a rather broad frequency window with $\omega > 160$ cm$^{-1}$. Interestingly there is a drop in the frequency resolved thermal conductivity at around 50 cm$^{-1}$, where low lying optical phonon modes are present. The effect of these low-lying optical modes on the thermal conductivity will be two-fold. First of all, they will open up scattering channels similar to the "rattler" of e.g. the skutterudites.[45] Secondly, the Grüneisen parameter is large for these modes indicating a large degree of anharmonicity. Additionally, these localized modes can be to a large degree be associated with the motions of Zn atoms (Fig. 2b). A detailed analysis of these vibrations reveal that they are mainly made up of the motions of Zn rotations around Se-Se dimers (as shown in Fig. 6a as an insert). In Fig 6a, the contribution of these vibrational motions to the phonon dispersions around 50 cm$^{-1}$ is shown. The large contribution means that the large anharmonicity at the low frequency optical phonon region (~50 cm$^{-1}$) comes mainly from the localized Zn-rotation modes in ZnSe$_2$.

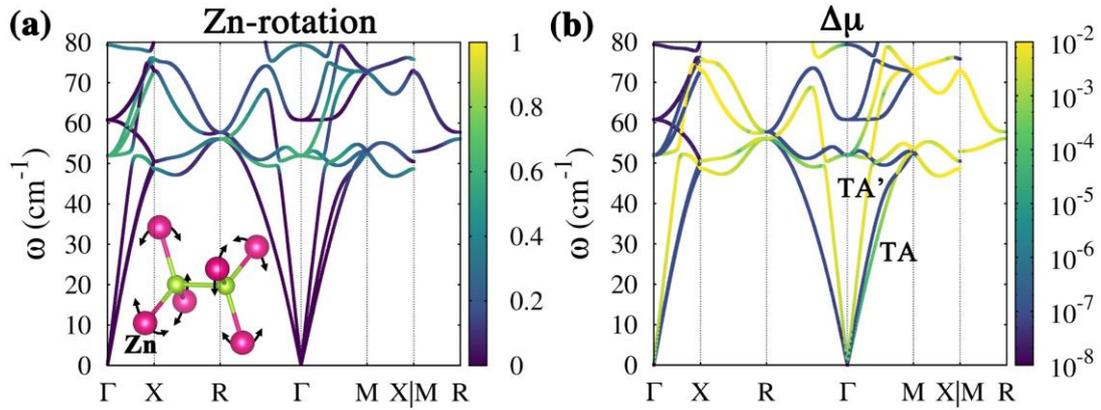

Fig 6 (a) The contribution of the motions of Zn rotations around the Se-Se dimer to the phonon dispersions around 50 cm$^{-1}$. The motions of Zn rotations around the Se-Se dimer vector is illustrated as an insert in (a). (b) The distribution of the changed pyrite parameter ($\Delta u$) at the acoustic phonons.

### B. Pyrite parameters

A further striking feature in Fig 5a, is that the Grüneisen parameters of certain acoustic bands are quite large. Consequently, it is also observed that the acoustic phonons make a surprising small overall contribution to the thermal conductivity, Fig. 5b, which will be an important factor in the overall low thermal conductivity. In particular, one of the transverse acoustic phonon branches (TA) along the $\overrightarrow{\Gamma M}$ direction has very large Grüneisen parameters ($\gamma > 4$), whereas another transverse acoustic phonon branch (TA') in the same direction has low $\gamma$ ($\gamma < 2$). This is a remarkable contrast that deserves further investigation, since whatever factor is behind the very different anharmonicities of these two branches may also help to better explain the anharmonicity character of the crystal as a whole.

As opposed to the localized Zn-rotation modes mentioned above, the TA phonon branches cannot be attributed to a specific type of localized vibration mode. Instead we will characterize them by the pyrite parameter ($u$) which, for a given pyrite-type crystal, determines the nonmetal atom positions.[46] When the whole crystal vibrates according to a particular phonon mode, the average pyrite parameter obviously does not change. However, for acoustic phonon vibrations, the parameter changes locally for relatively large volumes of the crystal. Hence, for a given unit cell, an acoustic phonon vibration which changes the pyrite parameter acts similarly to an expansion or contraction. This as we have already discussed, must trigger a large change to the TA phonon dispersion because of the significant anharmonicity and consequential large Grüneisen parameter. For the purposes of this work, we evaluate the linear sensitivity of the pyrite parameter ($\Delta u$) to each phonon mode by means of the following quantity:

$$\langle (u - u_0)^2 \rangle_H = \int_0^1 \{u[r_0 + H \times (\mathcal{R}\{v\}\cos(2\pi s) + \mathfrak{I}\{v\}\sin(2\pi s))] - u_0\}^2 \, ds$$

$$\Delta u = \lim_{H \to 0} \frac{\sqrt{\langle (u-u_0)^2 \rangle_H}}{H} \tag{9}$$

Where $\mathcal{R}\{v\}$ and $\mathfrak{I}\{v\}$ are the real and imaginary part of the phonon eigenvector ($v$), respectively, $u_0$ is the pyrite parameter of the pristine cell, and the value of $u(r)$ is calculated by fitting the coordinates to those of a pyrite unit cell in a least-squares sense. The calculated $\Delta u$ for the acoustic phonons are shown in Fig. 6b. From Fig. 6b, it becomes apparent that $\Delta u$ is significantly larger for the TA branch than for the TA' branch. This means that the large $\gamma$ at the acoustic phonon region can be explained in terms of the sensitivity of the pyrite parameter to different vibrational modes.

Through the above analysis, the strong anharmonicity mainly comes from two parts: the low-frequency optical phonons contributed by the motions of Zn rotations around the Se-Se dimer vector, and the transverse acoustic phonons with large Grüneisen parameter caused by the large change of the pyrite parameter. The strong anharmonicity of ZnSe$_2$ lead to the low lattice thermal conductivity [0.69 W/(m·K) at 300 K] at 0 GPa. Moreover, the previous studies[10,11] have shown that the weak interatomic interactions can contribute to the soft phonon dispersions and low lattice thermal conductivity. As mentioned above, comparing with 0 GPa, the interatomic bond lengths ($L_{\text{Zn-Se}}$ and $L_{\text{Se-Se}}$) are compressed at 10 GPa, resulting in an enhancement of the interatomic interactions and the higher-frequency phonon dispersions (as shown in Fig. 2c). Therefore, despite ZnSe$_2$ also has the large Grüneisen parameters at 10 GPa (Fig. S2b in Supplementary Materials), it is reasonable and easy to understand that $\kappa_l$ is increased to 1.98 W/(m·K) at 300 K.

## III. Electrical transport properties
### A. Band features

After calculating the lattice thermal conductivity, we investigate the electrical properties of ZnSe$_2$. The calculated electronic band structures at 0 GPa by PBE is shown in Fig. 7a. Since PBE is well known to underestimate the band gap of semiconductor[47], we also calculate the electronic band structures of ZnSe$_2$ at 0 GPa using HSE06 (Fig. 7b). From the two band structures (Fig. 7a and b), we find that the band gaps are 0.72 and 1.62 eV using PBE and HSE06, respectively. Fortunately, the band features around the Fermi level (valence bands and conduction bands) are similar in the two functional calculations. Therefore, we do not carry out the time-consuming HSE06 calculations with a highly dense k-point mesh for the electron BTE. Instead, we use the PBE electronic band structures and simply manually increase the band gap. At 10 GPa, the PBE calculated band gap is 0.61 eV (Fig. 7c). Assuming that the difference of the band gap (0.9 eV) using PBE and HSE06 does not change with pressure, we can set the band gap of ZnSe$_2$ at 10 GPa to ~ 1.51 eV.

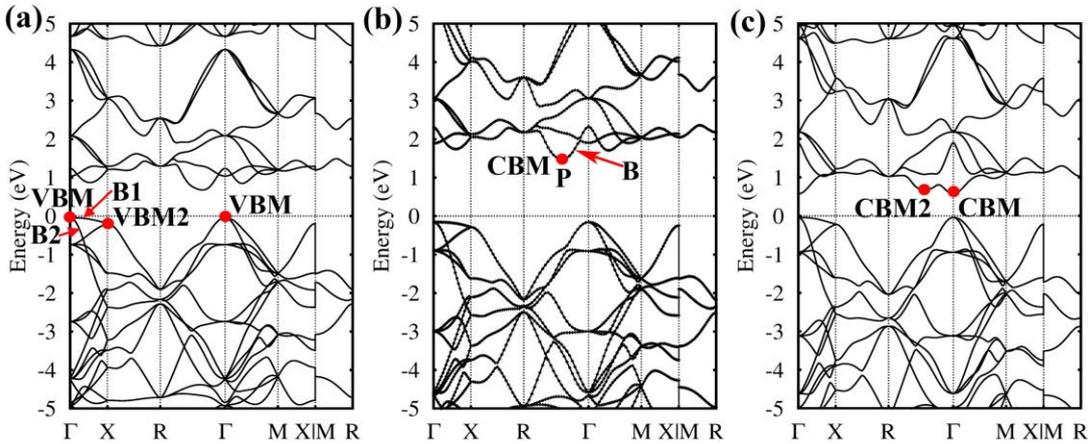

Figure 7. Calculated band structures of ZnSe$_2$ at 0 GPa for (a) PBE and (b) HSE approximations. The figure (c) is the PBE approximation calculated band structures of ZnSe$_2$ at 10 GPa.

The band behavior is further illustrated by the energy isosurfaces at 0 GPa as shown in Fig 8. The energy valleys of the conduction and valence bands in Fig. 8 relate to the corresponding energy band extrema in Fig. 7a at 0 GPa. From the band structures and energy isosurfaces of ZnSe$_2$ (Fig. 7a and 8a), we find the valence band maximum (VBM-Γ) and the secondary VBM (VBM-X) are located at Γ and X point, respectively. The energy difference between VBM-Γ and VBM-X is small (0.15 eV), which is comparable to that in the well-studied PbTe [48]. VBM-Γ and VBM-X both have a two-fold band-degeneracy and the small energy difference means both will fall within the width of the fermi distribution function and contribute to the electric transport properties. According to the shape of VBM-B1 in Fig. 8a, we notice that the p-type ZnSe$_2$ has a complex energy isosurface, which means that VBM-B1 possess both light and heavy carriers depending on direction [the corresponding carrier effective masses ($m^*_{\alpha\beta} = \hbar^2 \left[\frac{\partial^2 E(k)}{\partial k_\alpha \partial k_\beta}\right]^{-1}$) along different directions are list in Table. S1]. The light carriers can contribute the large σ, while the heavy carriers can facilitate the large $S$. Moreover, in a cubic structure, the symmetry degeneracy of VBM-X is six. This means that p-type ZnSe$_2$ has a large symmetry induced valley degeneracy. The complex energy isosurfaces suggest promising electrical properties of the p-type ZnSe$_2$. From the band structures of ZnSe$_2$ at different pressure (0 and 10 GPa, Fig. 7a and Fig. 7c), we notice that the valence band features are almost unaffected under pressure: The positions of VBM and VBM2 and the shapes of B1 and B2 do not change with pressure. However, the energy difference between VBM and VBM2 is slightly increased from 0.15 to 0.18 eV.

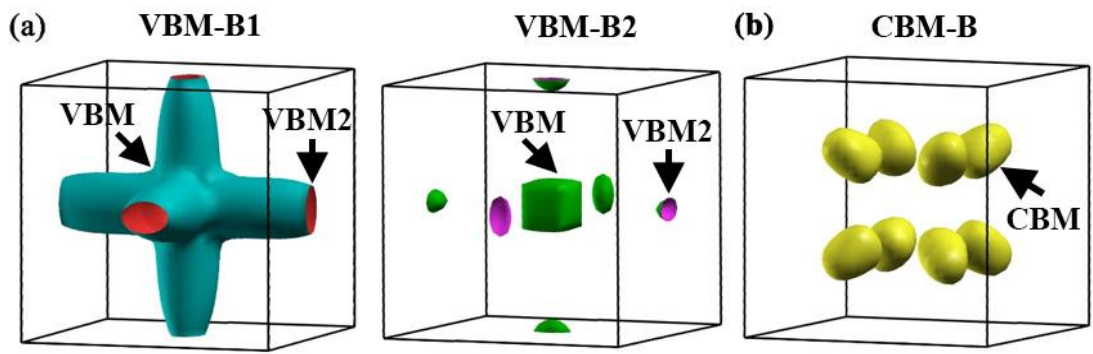

Figure 8. The energy isosurfaces at 0.2 eV low VBM and 0.15 eV above CBM for ZnSe$_2$ at 0 GPa.

For the conduction bands, the conduction band minimum (CBM) is located along

Γ-R direction (the P point in Fig. 7b) at 0 GPa. With increasing pressure to 10 GPa, the band features of the conduction bands are significantly changed: one band at the Γ point significantly lowers and becomes the CBM (CMB-Γ), while the band at the P point becomes the secondary conduction band minimum (CBM2-P). The CBM-P has a symmetry degeneracy is eight (Fig. 8b). The energy difference between CBM-Γ and CBM2-P at 10 GPa is small (0.12 eV). The shift of CBM results in the energy valley degeneracy for n-type $ZnSe_2$ is increased from 8 to 9 at 10 GPa. The shapes of CBM are near spherical, and the carriers of n-type $ZnSe_2$ along different directions have similar effective masses (the corresponding $m^*_{\alpha\beta}$ of CBM are listed in Table. S1). From Table. S1, we also find CBM-Γ and CBM2-P along different both have small $m^*_{\alpha\beta}$ and the mean $m^*_{\alpha\beta}$ is $0.42m_0$. The large degeneracy of the CBM and small carrier effective masses suggest that n-type $ZnSe_2$ also possess the promising electrical properties.

To further investigate the electronic properties of $ZnSe_2$ near Fermi level, we analyze the bonding characteristics at 0 GPa (Fig. 9a). The bonding characteristics are estimated using the LOBSTER package[49-52] to calculate the projected crystal orbital Hamilton population (COHP). The negative and positive COHP values indicate the bonding and antibonding character between atom pairs, respectively. Interestingly, both the bottom of the conduction bands ($E > E_f + E_g$) and the top of valence bands ($E < E_f$) are dominated by states with an antibonding character of Zn-Se and Se-Se. While it is somewhat unusual that the top of valence bands has an antibonding character, it is interesting to note that a similar feature has previously been found in materials with a complex conduction band edge dominated by several carrier pockets.[53] Moreover, through analyzing the projected band structures of $ZnSe_2$ (Fig. 9b), we find that the valence bands are mainly contributed by the electronic states of $Zn_d$ and $Se_p$, and the conduction bands are mainly contributed by the electronic states of $Zn_s$ and $Se_p$. For the valence bands, the weights of the same electronic state ($Zn_d$ or $Se_p$) to different bands are very similar. However, for the conduction bands, the first conduction band (CBM) is mainly contributed by the combination of $Zn_s$ and $Se_p$, while the second conduction band (CBM2) is solely contributed by $Se_p$. We thus infer that the electronic states of CBM at the P point (CBM-P) and CBM2 at the Γ point (CBM2-Γ) are contributed by the antibonding states of $Zn_s$-$Se_p$ and the antibonding states of $Se_p$-$Se_p$, respectively. The change of the relative bonding strength of Zn-Se and Se-Se will result the variation of the band energy difference between CBM-P and CBM2-Γ.

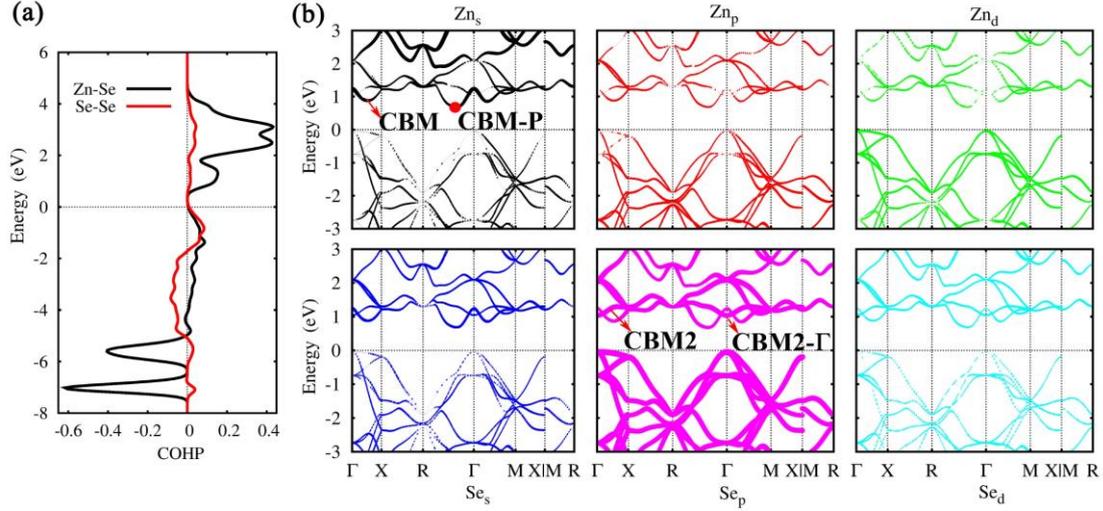

Figure 9, (a) The orbital-resolved projected crystal orbital Hamilton population (COHP) of ZnSe$_2$ at 0 GPa. (b) The projected decompose band structures of ZnSe$_2$ at 0 GPa. The thickness of the band feature represents the weights of the electronic state.

To better verify the change of the energy difference between CBM-P and CBM2-Γ ($\Delta E_{\Gamma P} = E_\Gamma - E_P$) with changing $\xi$, we calculate the change of the band structures of ZnSe$_2$ with $\eta_\xi$ [$\eta_\xi = (\xi - \xi_0)/\xi_0$] varying from +5% to -5%, as shown in Fig. S3 in Supplementary Materials. From Fig. S3, we can find that with decreasing $\eta_\xi$ (or the weakening Se-Se bonds), $\Delta E_{\Gamma P}$ decreases rapidly: from 0.82 to -0.03 eV. This means that the relative antibonding strength significantly influences the conduction band structures and the CBM position. Therefore, by tuning the relative antibonding strength, we could alignment the band energies of CBM2-Γ and CBM-P in ZnSe$_2$.

## B. Electrical transport properties

Typically, a large thermoelectric power factor is related to a band structure with a large valley degeneracy or the presence of both light and heavy effective mass carriers. While doping limits have been found to restrict the actual realization of good thermoelectric properties[54,55], the fact that both valence and conduction bands of ZnSe$_2$ exhibit a complex band-structures thus points to a considerable potential of this material for thermoelectric applications.

In order to calculate the electrical conductivities of ZnSe$_2$, we need to estimate the carrier relaxation time, $\tau$, Eq. (1), which requires the LA phonon velocities ($v_{LA}$), the single DOS effective masses ($m_b^*$) and the deformation potentials ($E_d$) of different valleys. Based on Eq. 3-7, we can calculate $v_{LA}$, $m_b^*$ and $E_d$ from the phonon dispersions (Eq. 3-4), the band structures (Eq. 5-6) and the energy change of the band extrema with volume (Eq.7), respectively. The values of $v_{LA}$, $m_b^*$ and $E_d$ at different valleys at 0 and 10 GPa are given in Table 1.

Table 1. Calculated longitudinal acoustic phonon velocity ($v_{LA}$), single DOS effective mass ($m_b^*$) and deformation potential ($E_d$) of different valleys at 0 and 10 GPa, respectively. $m_0$ is the mass of a free electron.

| P (GPa) | $v_{LA}$ (m/s) | $m_b^*$ ($m_0$) | | | | | $E_d$ (eV) | | | |
|---|---|---|---|---|---|---|---|---|---|---|
| | | VBM | | | CBM | | VBM | | CBM | |
| | | Γ | | X | Γ | P | Γ | X | Γ | P |
| | | B1 | B2 | | | | | | | |
| 0 | 3682 | 1.48 | 0.27 | 0.22 | - | 0.42 | -10 | -9.8 | - | -11.4 |
| 10 | 4392 | 1.24 | 0.22 | 0.20 | 0.34 | 0.75 | -11.6 | -11.4 | -7.5 | -10.6 |

From Table 1, we can find that when pressure is increased from 0 to 10 GPa, $v_{LA}$ is increased from 3684 to 4392 m/s due to the stiffer phonon dispersions (Fig. 2). Caused by the change of band structures, the calculated $m_b^*$ in VBM are slightly decreased but $m_b^*$ in CBM at P point is drastically increased from 0.42 to 0.75$m_0$, and the absolute value of $E_d$ ($|E_d|$) in VBM and CBM are slightly changed. From Eq.1-2, for p-type ZnSe$_2$, the increased $v_{LA}$ and decreased $m_b^*$ in VBM will lead to the increment of $\tau$. However, for n-type ZnSe$_2$, the increased $m_b^*$ in CBM at P point will lead to the decrement of $\tau$. The calculated $\tau$ of ZnSe$_2$ at 0 and 10 GPa are shown in Fig. 10a and Fig. 11a, respectively. From them, we find that $\tau$ is increased under pressure for p-type doping, but decreased for n-type doping.

Combining the calculated $\tau$ with the electronic BTE, we obtain the corresponding electrical transport properties ($S$, $\sigma$, and $PF$) of p-type and n-type ZnSe$_2$ at different temperatures and carrier concentrations at 0 and 10 GPa, as shown in Fig. 10b-d and Fig. 11b-d, respectively. Obviously, the increased carrier concentration deteriorates the Seebeck coefficient ($S$) but benefits the electrical conductivity ($\sigma$). Thus, the power factor ($PF = S^2\sigma$) reaches a maximum value at the carrier concentration around $10^{20}$ cm$^{-1}$ at 300 K. From Fig. 10 and 11, we find that: 1) both p-type (hole doping) or n-type (electron doping) ZnSe$_2$ have a high $PF$ [$PF_{max} > 1.0$ $m$W/(m·K$^2$)] at different pressures, which are comparable with the best TE material [1.01 $mW/(m\cdot K^2)$ for SnSe][8]. This means that $ZnSe_2$ do exhibit promising electrical properties. 2) The *PF* of p-type $ZnSe_2$ is increasing but n-type $ZnSe_2$ is decreasing under pressure. The pressure can be used to regulate the electrical transport properties of $ZnSe_2$.

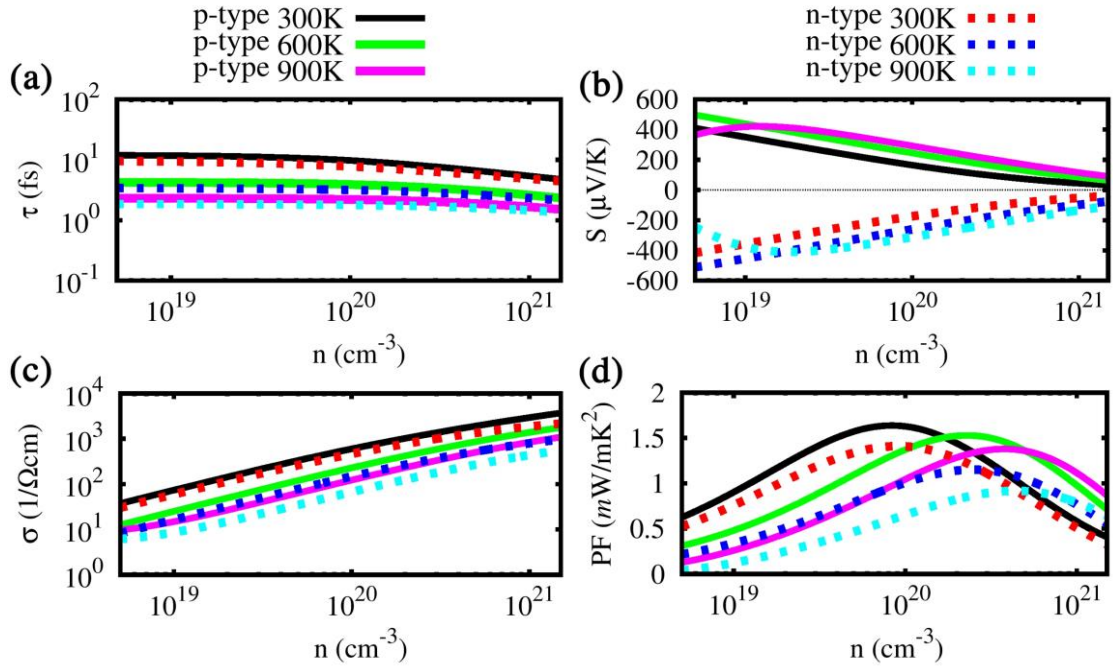

Figure 10. Calculated $\tau$ (a), *S* (b), $\sigma$ (c) and *PF* (d) of p-type and n-type $ZnSe_2$ as a function of the carrier concentration at 0 GPa, at 300, 600 and 900 K, respectively.

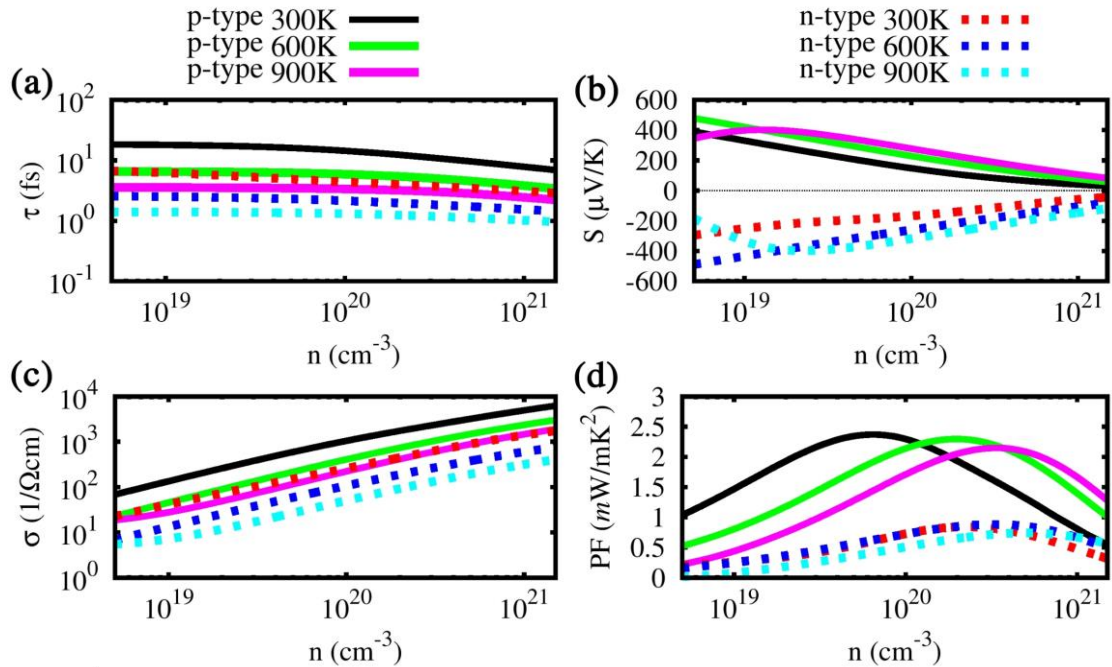

Figure 11. Calculated $\tau$ (a), $S$ (b), $\sigma$ (c) and $PF$ (d) of p-type and n-type ZnSe$_2$ as a function of the carrier concentration at 10 GPa, at 300, 600 and 900 K, respectively.

## IV. High ZT values

Combining the thermal ($\kappa_l$ and $\kappa_e$) and electrical ($PF$) properties of ZnSe$_2$, we can evaluate the $ZT$ values at different temperatures and carrier concentrations at 0 and 10 GPa, as shown in Fig. 12a and 12b, respectively. At 0 GPa, the maximum $ZT$ values of p-type and n-type ZnSe$_2$ can reach 2.21 and 1.87 at 900 K and at the carrier concentrations of $9.8 \times 10^{19}$ and $1.6 \times 10^{20}$ cm$^{-3}$, respectively. The high $ZT$ values indicate that ZnSe$_2$ is a promising thermoelectric material for both p-type and n-type doping. The comparative $ZT$ values of the p-type and n-type ZnSe$_2$ could be useful in designing two legs in the thermoelectric devices and also mean that, even if a doping limit should exist, at least either p- or n-type ZnSe$_2$ can be realized. With increasing pressure to 10 GPa, due to the increment of the electrical transport properties is smaller than the increment of the lattice thermal conductivity, the maximum $ZT$ values of the p-type ZnSe$_2$ is decreased to 1.51; moreover, due to the decrement of the electrical transport properties and the increment of the lattice thermal conductivity, the maximum $ZT$ values of the n-type ZnSe$_2$ is decreased to 0.77.

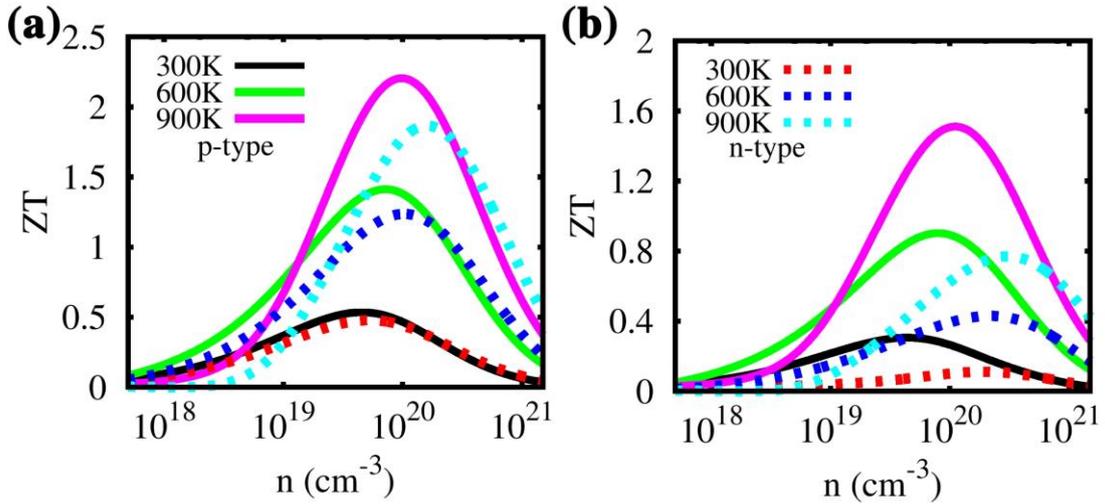

Figure 12. The $ZT$ values of ZnSe$_2$ as a function of temperature and carrier concentration for the p-type and n-type doping at 0 (a) and 10 (b) GPa, respectively.

## CONCLUSION

In this work, based on the phonon and electron Boltzmann transport theories, we have systematically studied the thermoelectric performances of ZnSe$_2$ at different pressures. These results show that ZnSe$_2$ has both low lattice thermal conductivities and promising electrical transport properties at different pressures. The low thermal conductivities come from the low-frequency optical phonons with a strong anharmonicity which are mainly contributed by the motions of Zn rotations around Se-Se dimers, and the acoustic phonons with a strong anharmonicity which are sensitive to the pyrite parameter. The promising electrical transport properties are contributed by the complex energy isosurfaces of both valence and conduction bands. At 0 GPa, the maximum *ZT* values of the p-type and n-type ZnSe$_2$ can reach 2.21 and 1.87, respectively. Our work shows that ZnSe$_2$ exhibits the excellent thermoelectric properties for both p-type and n-type doping. This work can provide a guidance and inspiration for the future experimental and theoretical works.

## ACKNOWLEDGMENTS

This work was supported by the National Natural Science Foundation of China, Grant No. 11774347 and 11474283. This work was supported by the China Scholarship Council.